# Insulator-to-Metal Transition via Magnetic Reconstruction at Oxide Interfaces

Zengxing Lu,[1, 2]† Jiatai Feng,[1, 2]† Xuan Zheng,[1, 2] You-guo Shi,[3] Run-Wei Li,[1, 4] Carmine Autieri,[5]* Mario Cuoco,[6]* Milan Radovic,[7]* and Zhiming Wang,[1, 2, 8]*

[1]*CAS Key Laboratory of Magnetic Materials and Devices, Ningbo Institute of Materials Technology and Engineering, Chinese Academy of Sciences, Ningbo 315201, China*
[2]*Zhejiang Province Key Laboratory of Magnetic Materials and Application Technology, Ningbo Institute of Materials Technology and Engineering, Chinese Academy of Sciences, Ningbo 315201, China*
[3]*Institute of Physics, Chinese Academy of Sciences, Beijing 100190, China*
[4]*Eastern Institute of Technology, Ningbo 315200, China*
[5]*International Research Centre Magtop, Institute of Physics, Polish Academy of Sciences, Aleja Lotników 32/46, PL-02668 Warsaw, Poland*
[6]*CNR-SPIN, UOS Salerno, I-84084 Fisciano (Salerno), Italy*
[7]*Photon Science Division, Paul Scherrer Institut, Villigen PSI, Switzerland*
[8]*Center of Materials Science and Optoelectronics Engineering, University of Chinese Academy of Sciences, Beijing 100049, China*

*Corresponding authors. E-mail:
autieri@magtop.ifpan.edu.pl
mario.cuoco@spin.cnr.it
milan.radovic@psi.ch
zhiming.wang@nimte.ac.cn

†These authors contributed equally to this work.

**Abstract:**

Ultrathin two-dimensional (2D) electronic systems at the interfaces of layered materials are highly desirable platforms for exploring of novel quantum phenomena and developing advanced device applications. Here, we investigate ultrathin heterostructures composed of $SrIrO_3$ (SIO) and $SrRuO_3$ (SRO) layers to uncover their emergent properties. Strikingly, despite the fact that both individual layers are antiferromagnetic insulators, the interfaced heterostructure exhibits emergent metallicity. Through transport measurements, magnetic characterization, and angle-resolved photoemission spectroscopy (ARPES), we analyze the underlying mechanisms governing this insulator-to-metal transition. Our findings reveal that the transition is driven by interface-induced magnetic reconstruction, which is further corroborated by density functional theory (DFT) calculations. The staggered Dzyaloshinskii-Moriya interaction at the SIO/SRO interface is identified as the key driving force for this spin reorganization, as it stabilizes ferromagnetism in the coupled antiferromagnetic insulating layers. These findings highlight the significant potential of engineering interfacial magnetic interactions as a powerful approach to generate and control emergent electronic properties, paving the way for novel functionalities that are unattainable in individual ultrathin films.

**Introduction**

Magnetic insulators are commonly characterized by antiferromagnetic order, which arises from Coulomb-driven Heisenberg superexchange. In contrast, insulating ferromagnetism (FM) is relatively rare in nature, as it necessitates a delicate balance of orbital degrees of freedom (*1*). Alternative mechanisms for ferromagnetism, such as double exchange (*2*) and Ruderman-Kittel-Kasuya-Yosida (RKKY) indirect exchange (*3-5*), are more effective in itinerant electronic systems and are often associated with metallic behavior. This dichotomy suggests that manipulating magnetic order through magnetic reconstruction or control of magnetic exchange, could potentially transform an insulator into a metal, and vice versa. While tuning electron density or introducing dopants to modify magnetic exchange has been explored, achieving metal-to-insulator transitions via genuine reconstruction of magnetic order remains a significant challenge in condensed matter physics. Here, we present a novel approach to control electronic properties and induce an insulator-to-metal transition (IMT) by engineering magnetic interactions at interfaces. Specifically, we demonstrate that magnetic coupling at the interface between two antiferromagnetic insulators can modify the magnetic state and induce metallic behavior.

Materials exhibiting strong spin-orbit coupling (SOC) and electron correlation are promising candidates for magnetic reconstruction-driven IMTs (*6-9*). The intricate interplay between SOC and electron correlation leads to a rich variety of electronic and magnetic ordering states that are highly sensitive to external stimuli, such as changes in temperature, pressure, or applied magnetic fields. These perturbations can trigger magnetic reconstructions and consequently induce IMTs. Among strongly correlated materials with significant SOC, transition metal oxides, particularly $SrIrO_3$ (SIO) and $SrRuO_3$ (SRO), have attracted considerable attention due to their unique dimensionality-dependent properties (*10-17*). In their bulk form, these oxides are metallic, but they undergo metal-to-insulator transitions in their ultrathin film limit, accompanied by modifications in their magnetic ordering (Fig. 1(A to D)). For instance, ultrathin SIO films exhibit a noncollinear antiferromagnetic insulating state (*13, 18, 19*), while ultrathin SRO films display a collinear antiferromagnetic insulating state (*20-27*). These thickness-dependent phase transitions, driven by the intricate interplay between SOC and electron correlation, render SIO and SRO ideal platforms for exploring magnetic reconstruction-induced IMTs in oxide heterostructures.

In this study, we demonstrate a novel route to achieve IMTs in oxide heterostructures via interfacial magnetic reconstruction. In heterostructures composed of ultrathin SRO and SIO layers, despite the individual layers display an antiferromagnetic insulating behavior, metallicity is shown to be accompanied by ferromagnetic properties. To demonstrate the electronic reconstruction, we systematically investigate SIO/SRO heterostructures with varying layer thicknesses through a combined experimental and theoretical approach. The emergence of a metallized ground state in the SIO top layer is evidenced by a significant spectral weight at the Fermi level, as revealed by *in-situ* angle-resolved photoemission spectroscopy (ARPES) using synchrotron radiation in the ultraviolet energy range. Theoretical analysis confirms that electronic states at the Fermi level emerge exclusively when ferromagnetism is induced by interface-driven magnetic reconstruction. In particular, our calculations reveal the pivotal role of the staggered Dzyaloshinskii-Moriya interaction (DMI) in reshaping the magnetic state within the SIO/SRO heterostructure, ultimately triggering the IMT.

**Results**
***A. Engineering Ferromagnetic Metallicity in SIO/SRO Heterostructures via Interfacial DMI***

To understand the mechanism of magnetic reconstruction at the SIO/SRO interface, we first consider the thickness-dependent properties of individual SIO and SRO films. Previous studies have shown that both materials undergo an IMT upon thickness reduction, accompanied by changes in their magnetic ordering (*14, 28, 29*). Using density functional theory (DFT) calculations and a model Hamiltonian approach, we investigate the ground state properties of ultrathin SIO and SRO films for a range of electron correlation strengths ($U_{Ir}$ and $U_{Ru}$). Our analysis support experimental findings, indicating that below a critical thickness of 3 unit cells (u.c.), both SRO and SIO films are insulating antiferromagnets (AFMs) with distinct magnetic structures. Specifically, SIO exhibits a non-collinear AFM (nc-AFM) order, whereas SRO displays a collinear AFM (c-AFM) arrangement, as shown in Fig. 1(A and C). These distinctions primarily arise from differences in SOC, magnetocrystalline anisotropy, and octahedral distortions inherent to each material.

The perovskite $ABO_3$ structure, with space group 62, accommodates a staggered DMI associated with to octahedral tilting and rotation (*30*). In bulk perovskites, where mirror symmetry is maintained, magnetic ordering is predominantly collinear. However, in SIO/SRO heterostructures, the staggered DMI (Fig. 1E) becomes significant at the interface due to the broken inversion and horizontal mirror symmetry. This interfacial DMI favors specific spin configurations with magnetic moments oriented along the out-of-plane and in-plane directions. The DMI is characterized by a vector $\vec{D}$ lying along a symmetry axis within the interface plane (*29, 31*), and it manifests as an antisymmetric magnetic coupling between spin moments $\vec{S}$ at Ru and Ir sites, described by $H_{DMI} = \vec{D} \cdot (\vec{S}_{Ru} \times \vec{S}_{Ir})$. This interaction induces a specific spin chirality in the plane perpendicular to $\vec{D}$, which can be either left- or right-handed depending on the staggered DMI. Consequently, the DMI plays a dual role in magnetic ordering at the interface. First, it transforms the noncollinear spin arrangement in SIO into a collinear configuration. Second, and more importantly, it drives a transition from AFM to FM order in the SRO electronic system. This AFM-to-FM transition is a direct consequence of the alternating DMI vector, which follows staggered octahedral distortions, and the broken inversion symmetry at the interface. The emergence of metallicity in the SIO/SRO heterostructure, as indicated by a finite density of states (DOS) at the Fermi level (Fig. 1F), is therefore a direct outcome of the DMI-driven magnetic reconstruction.

### B. Electrical and Magnetic Properties of SIO/SRO Heterostructures

To explore the emergent behavior stemming from interfacial coupling, we investigated the electrical transport and magnetic properties of SIO/SRO heterostructures. Figure 2A shows the temperature-dependent electrical resistivity $\rho(T)$ of individual 2-u.c. SIO and 2-u.c. SRO ultrathin films, alongside three representative heterostructures: 2-u.c. SIO/1-u.c. SRO, 1-u.c. SIO/2-u.c. SRO, and 2-u.c. SIO/2-u.c. SRO (hereafter denoted as 2SIO/1SRO, 1SIO/2SRO, and 2SIO/2SRO, respectively). These measurements were performed across a temperature range from 10 K to 300 K. While individual ultrathin films exhibit insulating behavior, all heterostructures demonstrate metallic temperature dependence, with conductivity increasing as the SRO layer thickness increases. Notably, kinks in the $\rho(T)$ curves are observed at temperatures ranging from 102 K and 123 K for the heterostructures, signifying a change in electron scattering rate associated with the paramagnetic-to-ferromagnetic phase transition of the SRO layer (*32, 33*).

To further probe the magnetic properties of the heterostructures, we performed magnetization versus magnetic field (*M - H*) hysteresis loop measurements (Fig. 2B). All heterostructures show clear ferromagnetic hysteresis loops, confirming the emergence of ferromagnetic order. The easy axis of magnetization is oriented along the surface normal direction. The saturated magnetic moments are estimated to be 0.6 $\mu_B$/Ru for 2SIO/1SRO and 1 $\mu_B$/Ru for both 1SIO/2SRO and 2SIO/2SRO heterostructures. These results collectively demonstrate the emergence of ferromagnetic ordering and metallic behavior in the SIO/SRO heterostructures, in stark contrast to the insulating antiferromagnetic nature of the individual ultrathin films.

### C. Probing Electronic Structure of Engineered Heterostructures

To gain a comprehensive understanding of the electronic structure and the emergence of metallic states in SIO/SRO heterostructures, we performed *in-situ* ARPES measurements. The valence band photoemission spectra (Fig. 2C) of individual 2-u.c. SIO and 2-u.c. SRO ultrathin films reveal a suppressed spectral weight near the Fermi level and the presence of an energy gap, consistent with their insulating nature. In contrast, the SIO/SRO heterostructures exhibit a finite DOS at the Fermi level (Fig. 2D), confirming their metallic character. This observation provides direct evidence for the electronic structure reconstruction induced by the heterogeneous integration of ultrathin SIO and SRO layers. To further elucidate the emergent metallic states, we measured the Fermi surface and band dispersions of the heterostructures using ARPES with a circularly right-polarized synchrotron light ($hv = 82$ eV), as shown in Fig. 3. The Fermi surface maps (Fig. 3 (A.i to C.i)) reveal that the electronic band crosses the Fermi level and form a well-defined Fermi surface in all three heterostructures, with a hole-like pocket centered at the X point of the Brillouin zone. The band dispersion along the high-symmetry X-Γ-X direction (Fig. 3 (A.ii to C.ii)) exhibits a clear quasiparticle peak at the Fermi level in the extracted energy distribution curves (EDCs) at the X point (Fig. 3 (A.iii to C.iii)), which is a hallmark of metallic behavior. The relative weight of the coherent peak increases from 2SIO/1SRO to 1SIO/2SRO and 2SIO/2SRO, indicating an enhancement of the metallic character with increasing SRO thickness, consistent with the electrical transport measurements.

To quantitatively analyze the band dispersions, we extracted the momentum distribution curves (MDCs) and fitted them with a parabolic function (Fig. 3D). The effective electron mass ($m^*$) decreases from 0.57 $m_e$ for the 2SIO/2SRO heterostructure to 0.5 $m_e$ and 0.41 $m_e$ for the 1SIO/2SRO and 2SIO/1SRO heterostructures, respectively, further confirming the enhanced metallicity with increased total thickness. These results demonstrate that the electronic structure of the SIO/SRO heterostructures can be effectively tuned by controlling the relative thicknesses of the SIO and SRO layers. The direct observation of a finite DOS at the Fermi level, the presence of a well-defined Fermi surface, and the evolution of effective electron mass with heterostructure thickness provide compelling evidence for the interfacial coupling-driven electronic reconstruction in these engineered oxide heterostructures.

### *D. DMI-Driven Magnetic Reconstruction and its Impact on Electronic Structure*

To gain deeper insight into the IMT mechanism in SIO/SRO heterostructures, we performed first-principles simulations based on local spin density approximations and constructed an effective model Hamiltonian. This model incorporates nearest-neighbor magnetic exchange between Ru and Ir spin moments at the SIO/SRO interface. Our analysis reveals that the staggered DMI at the interface plays a crucial role in the magnetic reconstruction process. Specifically, in our unit cell, which contains two spins per atomic species, we focus on the DMI between Ir and Ru spins, as depicted in Fig. 1E. DMI tends to produce magnetic configurations where Ru and Ir spins are rotated by 90 degrees with respect to each other. The simplest spin model with a staggered DMI as the dominant interaction between two layers predicts one layer with collinear FM order and another with c-AFM order. Therefore, the staggered nature of the DMI favors a noncollinear magnetic phase between SIO and SRO but collinear phases within the SIO and SRO layers themselves, which are more metallic compared to the nc-AFM order in pure SIO and AFM order in pure SRO. To understand the impact of magnetic reconstruction on electronic structure, we compared the DOSs for three-layer systems with different structural configurations in their respective magnetic ground states: c-AFM SRO, nc-AFM SIO, and two types of

SIO/SRO interfaces with c-AFM SIO and FM SRO.

Orbital-resolved DOSs are presented in Fig. 4 for individual insulating 2-u.c. SIO and SRO films, and metallic three-layer heterostructures. A clear gap between the valence band and conduction band in 2SIO and 2SRO confirms the insulating nature at the critical thickness for the IMT. However, the pronounced spectral weight of the DOS at the Fermi level in the 2SIO/1SRO heterostructure indicates the emergence of a metallic state. The DOS calculations for the 1SIO/2SRO heterostructure exhibit a similar behavior to that of the 2SIO/1SRO system. Additionally, they reveal a DOS spectrum that continuously crosses the Fermi level, further substantiating the emergence of a metallic state. Moreover, the calculations indicate higher conductivity in the 1SIO/2SRO heterostructure compared to 2SIO/1SRO, which is consistent with our experimental observations.

**Discussion**

Our study showcases a novel approach to induce IMT in oxide heterostructures through interfacial magnetic reconstruction. By engineering magnetic interactions at the SIO/SRO interface, we have demonstrated the emergence of a ferromagnetic metallic state from two initially insulating antiferromagnetic oxides. This magnetic reconstruction-driven IMT is fundamentally different from conventional metallic states observed at interfaces of bulk band insulators, such as the well-studied $LaAlO_3$/$SrTiO_3$ (LAO/STO) system (*34, 35*). In LAO/STO interfaces, a highly conductive two-dimensional electron gas (2DEG) forms due to electronic reconstruction caused by polar discontinuity (*36, 37*). In contrast, the emergent metallicity in SIO/SRO heterostructures arises from a delicate interplay among various magnetic interactions, including Ru-Ir coupling, magnetocrystalline anisotropy, and the staggered DMI. Our first-principles calculations and model Hamiltonian analysis indicate that these interactions compete to stabilize the magnetic ground state, with the staggered DMI favoring collinear FM order in SRO and c-AFM order in SIO. Anisotropic Ru-Ir coupling, exhibiting AFM coupling along the out-of-plane direction and FM coupling along the in-plane direction, competes with the magnetocrystalline anisotropy of Ir ions, which tends to align spins along the out-of-plane direction. In addition, DMI promotes a chiral spin texture perpendicular to spin orientations favored by Ru-Ir coupling and magnetocrystalline anisotropy. The emergence of this chirality is attributed to the broken mirror and rotational symmetries. The presence of a chiral spin structure with a ferromagnetic background has been previously observed in $SrRuO_3$ thin films (*38*), and we anticipate a similar physical origin in our system. The equilibrium among these competing interactions ultimately dictates the magnetic ground state and the resulting electronic structure of the system. In SIO/SRO heterostructures, DMI dominates over other interactions, thereby driving magnetic reconstruction and the emergence of metallicity.

The emergence of DMI in SIO/SRO heterostructures is attributed to the strong SOC

and the broken inversion symmetry at the interface, which facilitates magnetic reconstruction. The DMI-driven magnetic reconstruction mechanism is distinct from conventional scenarios where modifications in magnetic order are primarily mediated by exchange interactions, often influenced by doping and strain (*39*). The staggered nature of the DMI, originating from oxygen octahedral distortions at the SIO/SRO interface, plays a pivotal role in this magnetic reconstruction. Specifically, it converts the nc-AFM order in the SIO layer into a c-AFM configuration between Ir moments, while concurrently inducing a transition from an antiferromagnetic to a ferromagnetic state in the SRO layer. This intimate link between the staggered DMI and octahedral distortions at the interface underscores the critical importance of spin-charge-lattice coupling in governing emergent properties in oxide heterostructures (*40-42*). Consequently, the strength and sign of the DMI can be tuned by various factors, including interface engineering and application of external electric and magnetic fields (*43*). This tunability enables the dynamic control of chiral magnetic states, offering promising avenues for designing and optimizing novel chiral spintronic devices (*44*). Furthermore, the ability to manipulate and switch between different chiral magnetic states in SIO/SRO heterostructures by modulating the DMI could pave the way for the development of advanced spintronic functionalities, such as high-density memory storage and energy-efficient information processing (*45-47*).

In summary, our work demonstrates a novel mechanism for inducing IMT in oxide heterostructures through the precise engineering of interfacial magnetic interactions. The DMI-driven magnetic reconstruction in SIO/SRO heterostructures represents a significant step forward in our understanding of the intricate interplay between magnetic and electronic degrees of freedom in these systems, and it opens new avenues for the rational design of functional oxide interfaces. While challenges and opportunities for further exploration remain, we believe that DMI-driven magnetic reconstruction will emerge as an essential tool in the pursuit of novel functionalities and device concepts based on oxide heterostructures.

**Materials and Methods**
*A. Fabrication of Heterostructures*

SrIrO$_3$/SrRuO$_3$ (SIO/SRO) heterostructures were grown as bilayers, including 1 u.c.-SIO/2u.c.-SRO (1SIO/2SRO), 2 u.c.-SIO/1 u.c.-SRO (2SIO/1SRO), and 2 u.c.-SIO/2 u.c.-SRO (2SIO/2SRO). All heterostructures were epitaxially grown on TiO$_2$-terminated [001]-SrTiO$_3$ (STO) substrates via pulsed laser deposition (PLD) assisted by reflection high-energy electron diffraction (RHEED). Polycrystalline SIO and SRO targets were ablated using a KrF excimer laser and Nd:YAG solid-state laser. SRO and SIO layers were successively grown at 650 °C in 0.15 mbar oxygen pressure. The laser fluence was 1.5 J/cm$^2$ with a repetition rate of 2 Hz. Layer-by-layer growth was ensured by monitoring *in-situ* RHEED patterns and oscillations throughout deposition. Following deposition, heterostructures were annealed at 650 °C for 30 min in 1 mbar oxygen pressure to promote full oxidation and then cooled to room temperature at a cooling rate of 10 °C/min while maintaining oxygen pressure. For magnetic and electric transport measurements, a 15-u.c. STO capping layer was deposited on the grown ultrathin films prior to annealing to prevent degradation under ambient conditions.

## B. Magnetic and Transport Measurements

Magnetic measurements were performed *ex-situ* at 10 K using a Quantum Design superconducting quantum interference device (SQUID). Transport measurements were conducted using a physical property measurement system (PPMS) in a van der Pauw configuration.

## C. ARPES Measurements

*In-situ* ARPES measurements were performed at the SIS beamline of the Swiss Light Source, Paul Scherrer Institute, utilizing a combined growth and spectroscopy setup. ARPES data were acquired at 10 K using circularly right-polarized synchrotron light with an energy of 82 eV under ultrahigh vacuum ($< 1 \times 10^{-10}$ mbar). Low-energy electron diffraction (LEED) patterns were obtained after ARPES measurements.

## D. First-Principles Calculations

First-principles density functional theory (DFT) calculations were performed using the planewave VASP package (*48, 49*) within the local spin density approximation (LSDA) and employing the Perdew-Zunger parametrization (*50*) of Ceperley-Alder data (*51*) for the exchange-correlation functional. Core-valence electron interactions were described using the projector augmented wave (PAW) method (*52*). On-site Coulomb interaction $U$ and Hund's exchange $J_H$ were included for Ru, Ir, and Ti 3$d$ orbitals within the LSDA+U approach, using a rotationally invariant scheme with $U = 5$ eV and $J_H = 0.15\ U$ (*53*). To simulate experimental conditions, we modeled STO-SRO, STO-SIO, and STO-SRO-SIO heterostructures with SrO termination using a slab geometry with a 20 Å vacuum layer. The STO thickness was fixed at 3 unit cells (u.c.), while SRO-SIO layers were varied with 2 to 4 u.c.. A plane wave cutoff of 480 eV was used. Brillouin zone integration was performed using $8 \times 8 \times 1$ and $12 \times 12 \times 1$ $k$-point grids for geometric relaxation and density of states/total energy/charge/magnetization calculations, respectively. Spin-orbit coupling (SOC) was included in all calculations except for geometric relaxation, which was performed in the non-relativistic limit for computational efficiency. Here, we study properties in the fully distorted phase with the in-plane lattice constant fixed to that of STO (3.905 Å). Internal degrees of freedom were optimized by minimizing total energy to below $3 \times 10^{-5}$ eV and Hellmann-Feynman forces to below 20 meV/Å.

**Acknowledgments**
*Funding:* Z.W. was supported by the National Key Research and Development Program of China (Grants Nos. 2024YFA1410200, 2019YFA0307800), the National Natural Science Foundation of China (Grants Nos. 12174406, U24A6001, 52127803), the Chinese Academy of Sciences Project for Young Scientists in Basic Research (No.YSBR-109), the Key Research Program of Frontier Sciences, Chinese Academy of Sciences (Grant No. ZDBS-LY-SLH008), the Ningbo Key Scientific and Technological Project (Grant No. 2022Z094), the Ningbo Natural Science Foundation (No. 2023J411). C.A. was supported by the Foundation for Polish Science project "MagTop" no. FENG.02.01-IP.05-0028/23 co-financed by the European Union from the funds of Priority 2 of the European Funds for a Smart Economy Program 2021-2027 (FENG). C.A. acknowledges access to the computing facilities of the Interdisciplinary Centre of Modelling at the University of Warsaw, Grant No. GB84-0, GB84-1 and GB84-7. C.A. acknowledges the CINECA award under the ISCRA initiative IsC85 "TOPMOST" and IsC93 "RATIO" Grants, for the availability of high-performance computing resources and support. The authors acknowledge beamtime on the SIS beamline of the Swiss Light Source and Dreamline of the Shanghai Synchrotron Radiation Facility. M.C. acknowledges support from the EU's Horizon 2020 research and innovation program under Grant Agreement No. 964398 (SU-PERGATE) and from PNRR MUR project PE0000023-NQSTI.

*Author contributions:* M.R. and Z.W. conceived and devised the project with input from all authors. Z.L. performed transport measurements. J.F. and X.Z. performed ARPES measurements. C.A. performed computational analysis and developed theoretical models. M.R., Z.W., C. A., and M.C. wrote the manuscript, with contributions from all authors.

**Competing interests:** Authors declare that they have no competing interests.

**Data and materials availability:** All data are available in the main text.

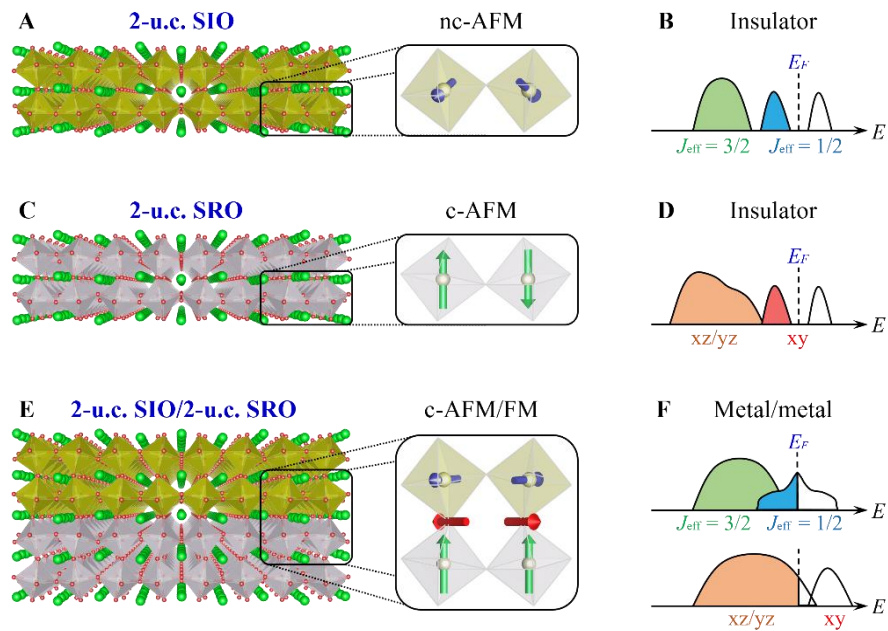

**Fig. 1. Schematic mechanism of DMI-induced insulator-to-metal transition in SIO/SRO heterostructures.** Magnetic structures and density of states (DOS) of (**A** and **B**) 2-u.c. SIO, (**C** and **D**) 2-u.c. SRO, and (**E** and **F**) SIO/SRO heterostructures. Arrows represent spin moments, with blue and green colors denoting Ir and Ru moments, respectively. Ultrathin SIO and SRO are noncollinear antiferromagnetic (nc-AFM) and collinear antiferromagnetic (c-AFM) insulators, respectively, as indicated by gaps in the DOS. In SIO/SRO heterostructures, staggered DMI on apical oxygens (red arrows with opposite directions) induces magnetic reconstruction, leading to a collinear antiferromagnetic (c-AFM) state in the SIO layer and a FM state in the SRO layer. Emergent metallicity in the heterostructure, evidenced by finite DOS at the Fermi level, is a direct consequence of DMI-driven magnetic reconstruction at the SIO/SRO interface. Orbital contributions to the DOS are indicated.

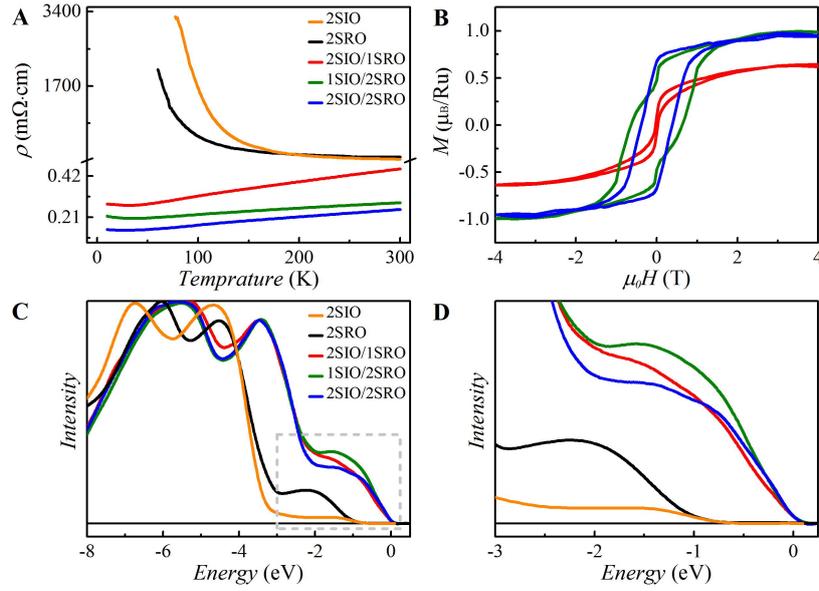

**Fig. 2. Electrical and magnetic properties of SIO/SRO heterostructures. (A)** Temperature-dependent electric resistivity $\rho(T)$ and **(B)** magnetization hysteresis loops $M(\mu_0 H)$. **(C)** Valence band spectra of ultrathin bilayers compared with 2-u.c. SRO and 2-u.c. SIO monolayers. **(D)** Magnified plots of spectra curves in **(C)** near the Fermi level.

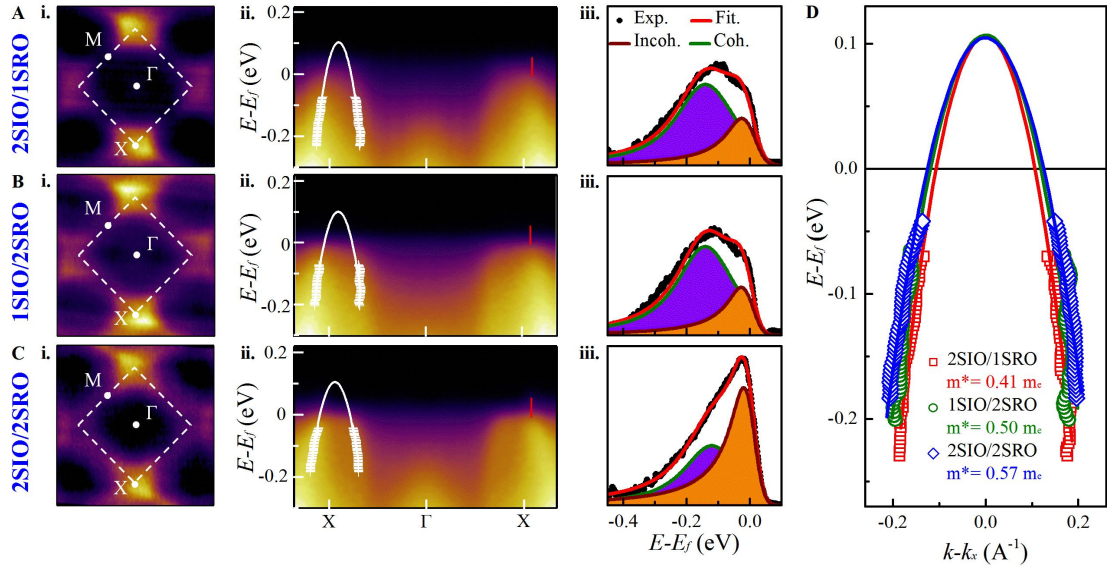

**Fig. 3. Electronic structure of SIO/SRO heterostructures probed by ARPES.** Fermi surface, dispersion along X-Γ-X (white dots and plots indicate locations of EDCs), and coherent/incoherent peaks derived from EDCs around X points (marked by short red lines in dispersions) for **(A)** 1SIO/2SRO, **(B)** 2SIO/1SRO, and **(C)** 2SIO/2SRO. Exp., Fit, Incoh., and Coh. are abbreviations for experiment, fit, incoherent, and coherent parts of data, respectively. **(D)** Extracted dispersion along X−Γ for films, with parabolic fits. Electronic dispersion is plotted relative to band maximum. $m^*$ and $m_e$ denote electron effective mass and bare electron mass, respectively.

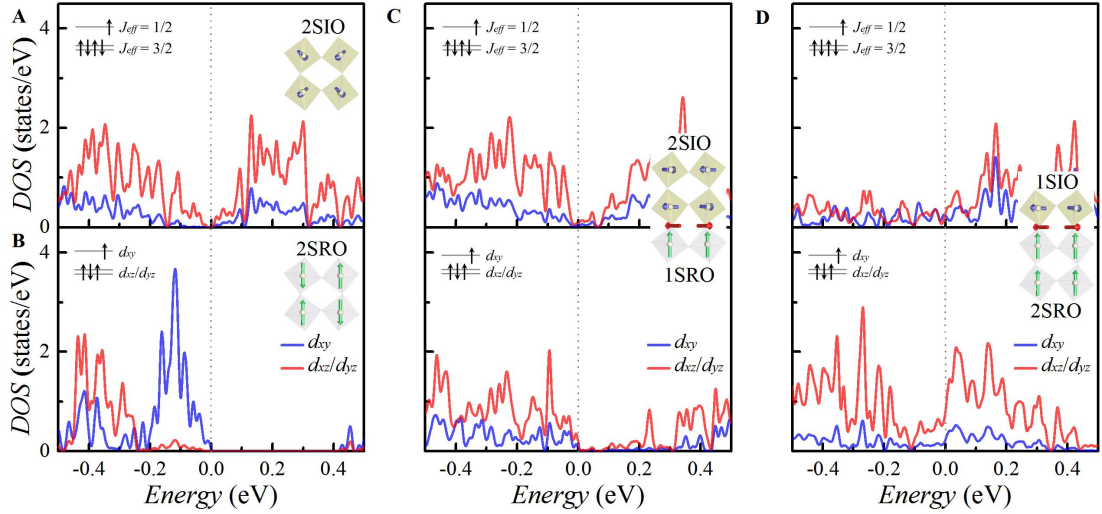

**Fig. 4. Orbital-resolved density of states (DOSs) of individual thin films and heterostructures.** DOS comprised of $d_{xy}$ (blue) and degenerate $d_{xz}/d_{yz}$ (red) orbitals are depicted for the films of **(A)** 2SIO and **(B)** 2SRO, and the heterostructures of **(C)** 2SIO/1SRO and **(D)** 1SIO/2SRO. Magnetic order and electronic configurations are indicated in the insets, where SIO and SRO crystal structures are drawn in dark yellow and grey-white. Spin ordering for Ir and Ru are represented by blue and green arrow, respectively. At the Ir-Ru interface, the staggered DMI (red arrow) stabilizes non-collinear spin orientation between Ir and Ru.